\documentclass[showpacs, preprintnumbers, nofootinbib, aps, prd, superscriptaddress,10pt, showkeys, notitlepage, twocolumn]{revtex4-1}

\usepackage{graphicx,amssymb,amsmath,amsthm,amsfonts,epsfig}

\usepackage[linktocpage,breaklinks]{hyperref}
\usepackage[usenames,dvipsnames]{color}
\usepackage{epstopdf}
\usepackage{aas_macros}
\usepackage{pifont}
\definecolor{darkred}{rgb}{0.5,0,0}
\definecolor{darkgreen}{rgb}{0,0.5,0}
\definecolor{darkblue}{rgb}{0,0,0.5}
\definecolor{prussian}{rgb}{0.0, 0.19, 0.33}
\definecolor{richelectricblue}{rgb}{0.03, 0.57, 0.82}
\definecolor{teal}{rgb}{0.0, 0.5, 0.5}
\definecolor{mediumseagreen}{rgb}{0.24, 0.7, 0.44}
\definecolor{lust}{rgb}{0.9, 0.13, 0.13}
\definecolor{ballblue}{rgb}{0.13, 0.67, 0.8}
\definecolor{darkcyan}{rgb}{0.0, 0.55, 0.55}
\definecolor{mountainmeadow}{rgb}{0.19, 0.73, 0.56}
\definecolor{palecarmine}{rgb}{0.69, 0.25, 0.21}
\definecolor{richcarmine}{rgb}{0.84, 0.0, 0.25}
\definecolor{tangelo}{rgb}{0.98, 0.3, 0.0}
\definecolor{venetian}{rgb}{0.784,0.031,0.082}
\definecolor{bdfrance}{rgb}{0.192,0.549,0.906}

\hypersetup{colorlinks=true, citecolor=venetian,
linkcolor=bdfrance, urlcolor=lust}
\usepackage{amsmath,amssymb}
\usepackage{tensor}
\usepackage{mathtools}
\usepackage{amsbsy}
\usepackage{bm}
\usepackage{float}

\newcommand{\be}{\begin{equation}}
\newcommand{\ee}{\end{equation}}
\newcommand{\bea}{\begin{eqnarray}}
\newcommand{\eea}{\end{eqnarray}}

\newcommand{\cS}{{\cal S}}

\newcommand{\rph}{{\rm ph}}

\begin{document}

\title{Can supermassive black hole shadows test the Kerr metric?}

\begin{abstract}

The unprecedented image of the M87* supermassive black hole has sparked some controversy over its usefulness as a test 
of the general relativistic Kerr metric. The criticism is mainly related to the black hole's quasi-circular shadow and advocates 
that its radius depends not only on the black hole's true spacetime properties but also on the poorly known physics of the 
illuminating accretion flow. In this paper we take a sober view of the problem and argue that our ability to probe gravity with a 
black hole shadow is only partially impaired by the matter degrees of freedom and the number of non-Kerr parameters used 
in the model. As we show here, a more intriguing situation arises from the mass scaling of the dimensional 
coupling constants that typically appear in non-GR theories of gravity. Existing limits from gravitational wave observations imply 
that supermassive systems like the M87* black hole would suffer a suppression of all non-GR deviation parameters in their metric, 
making the spacetime and the produced shadow virtually Kerr. Therefore, a supermassive black hole shadow is likely to 
probe only those extensions of General Relativity which are endowed with dimensionless coupling constants 
or other special cases with a screening mechanism for black holes or certain types of spontaneous scalarisation.

\end{abstract}

\author{Kostas Glampedakis}
\email{kostas@um.es}
\affiliation{Departamento de F\'isica, Universidad de Murcia, Murcia, E-30100, Spain}
\affiliation{Theoretical Astrophysics, University of T\"ubingen, Auf der Morgenstelle 10, T\"ubingen, D-72076, Germany}

\author{George Pappas}
\email{gpappas@auth.gr}
\affiliation{Department of Physics, Aristotle University of Thessaloniki, Thessaloniki 54124, Greece}

\date{\today}
 
\maketitle


\section{Introduction}
The centenary of the Eddington-Dyson 1919 observation of light deflection by the sun \cite{Dyson1920RSPTA} was marked by another 
important milestone in gravitational physics, the release of the direct image of the supermassive black hole in the nucleus of the M87 galaxy 
by the Event Horizon Telescope (EHT) collaboration \cite{EHT2019ApJ.I,EHT2019ApJ.V,EHT2019ApJ.VI}. This millimeter-band radio image 
of unprecedented angular resolution, itself an example of extreme light deflection, has provided direct quantitative evidence of the presence of 
supermassive  black holes in galactic centers and has shed some light in the inner workings of active galactic nuclei. A second image, that of our 
galactic SgrA* supermassive black hole, is scheduled to be released by the EHT collaboration in the near future. 

From the point of view of fundamental physics, a key element of an image like that of the M87* black hole is the geometric 
shape of the shadow seen by an asymptotic observer, as superimposed in the brighter background of the luminous matter surrounding the
black hole. By its very nature, and in contrast to the observation of gravitational waves from compact binary systems, the generation and observation 
of a black hole image is an `experiment' on the geodesic motion of photons emitted by the accretion flow, and therefore probes the geometry rather 
than the dynamical properties of the classical spacetime. 

One of the key motivations behind the conception of EHT was to use the shadow as evidence for the existence of
black holes and as a probe of General Relativity (GR) (see e.g. \cite{Broderick2014ApJ} for a review). This exciting possibility has spawn a significant amount 
of work over the last decade or so, mostly focused on the calculation of shadows of non-Kerr black holes beyond GR~\cite{Johannsen2010ApJ, 
Johannsen2013ApJ, Cunha2015PhRvL,Cunha2016IJMPD, Cunha2016PhRvD, Cunha2017PhysRevD, Cunha2017PhysRevLett, Cunha2018GReGr,Medeiros:2019cde} 
but also on improving our understanding of the image produced by garden-variety Kerr black holes \cite{Gralla_etal2019, Gralla2020, Gralla_Lupsasca2020}.

Throughout this paper we use relativistic units $G=c=1$.


\section{Probing gravity with black hole shadows} 
\begin{figure}[htb!]
\begin{center}
\includegraphics[width=0.5\textwidth]{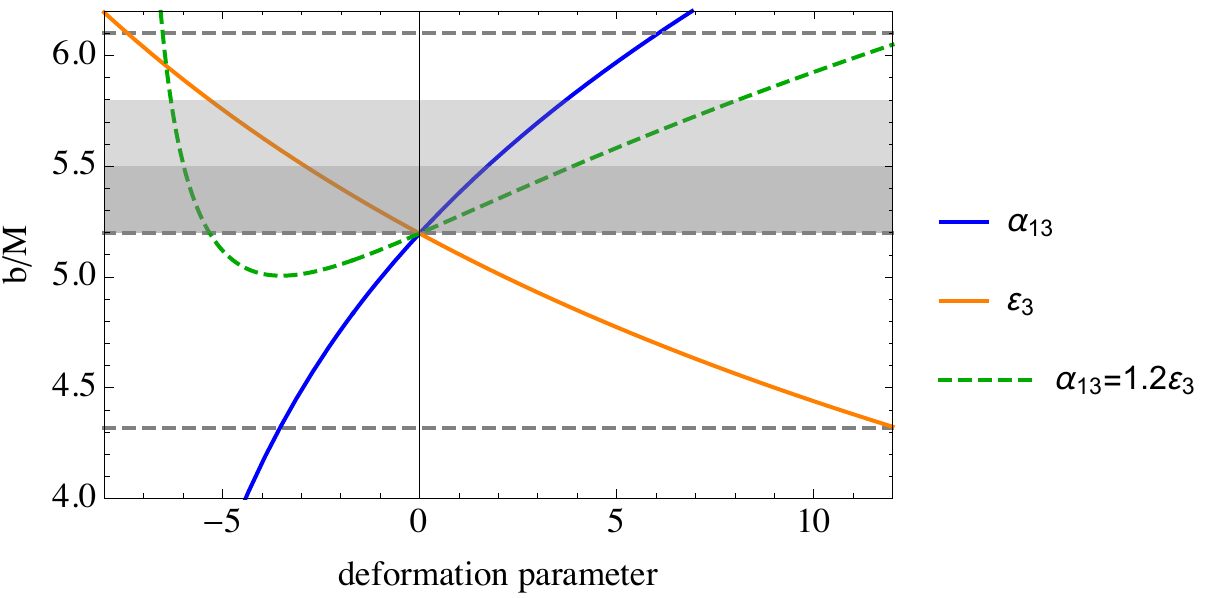}
\end{center}
\caption{The impact parameter/shadow radius $b$ associated with the photon ring of a non-rotating  black hole in the J-metric 
as a function of the deformation parameters $\alpha_{13}, \varepsilon_3$.  The middle dashed line indicates the GR value 
$b_{\rm GR}=3\sqrt{3}M$ while the upper/lower dashed lines mark the $\approx 17\%$ observational error in the image of M87*
\cite{Psaltis_etal2020}. The shaded region represents the range in the GR value of $b$ due to the variation in the geometry of the 
accretion flow (spherical, simple thick disk, or more realistic GR-MHD accretion) [light shade: as estimated 
in \cite{Gralla_etal2019}; 
dark shade: as estimated 
in \cite{Bronzwaer:2020vix}]. 
The dashed curve represents the case where the deformation parameters are correlated as 
$\alpha_{13} = 1.2 \varepsilon_3$ (with the horizontal axis measuring $\varepsilon_3$), 
resulting in a shadow radius much closer to $b_{\rm GR}$.}
\label{fig:bJ_shadow}
\end{figure}

In a recent paper, Psaltis et al.~\cite{Psaltis_etal2020} used the physical shape of the M87* black hole shadow as a test of GR. 
A prerequisite for this type of test is the independent knowledge of the black hole's mass, that is, the system's intrinsic 
yardstick. In the case of M87* the mass has been estimated to be $M =(6.6 \pm 0.4)\times 10^9 M_{\odot}$ by stellar kinematics 
with the assumption of a distance of $17.9 \textrm{Mpc}$ \cite{Geb2011ApJ}. The deviation from GR was modelled with the help of 
the Johannsen metric~\cite{Johannsen:2013pca} (hereafter `J-metric') which is a theory-agnostic deformation of the Kerr metric.
To lowest order, the deformation is encapsulated in the four constant \emph{dimensionless} parameters 
$\{\alpha_{22}, \alpha_{13}, \alpha_{52}, \varepsilon_3 \}$. The J-metric is Kerr-like in the sense that it is separable (thus admitting a third 
constant of motion like the Carter constant),  admits spherical photon orbits \cite{Glampedakis_Pappas2019}, and is endowed with a spherical 
event horizon with the same radius $r_{+} =M + \sqrt{M^2-a^2}$ as a Kerr black hole of the same mass $M$ and spin $a$.
Therefore, it is not too surprising that  black holes in the J-metric cast Kerr-like shadows \cite{Johannsen2013ApJ, Medeiros:2019cde}, namely,  
shadows that have a nearly constant circular radius unless the spin parameter $a$ lies close to its maximum allowed value. Based on this property, 
one can focus on a non-rotating system for which the shadow is exactly circular with its radius given by the impact parameter $b$ associated with 
the unstable circular photon orbit, i.e. the black hole's photon ring. The photon ring radius $r_\rph$ and the impact parameter depend on the single 
metric component~\cite{Cardoso_etal2009, Psaltis_etal2020}
\be
g_{tt} (r) = - \left (1-\frac{2M}{r}  \right )  \frac{ \left (1+ \varepsilon_3 M^3/r^3 \right ) }{\left (1+ \alpha_{13} M^3/r^3 \right )^2}, 
\ee
and are given by
\be
b = \frac{r_\rph}{\sqrt{-g_{tt}(r_\rph)}}, \qquad r_\rph \frac{dg_{tt}}{dr} (r_\rph) = 2 g_{tt} (r_\rph).
\ee
The aforementioned strategy was adopted in Ref.~\cite{Psaltis_etal2020} and the results are summarised in Fig.~\ref{fig:bJ_shadow} where 
we show $b$ for a non-rotating black hole in the J-metric as a function of the deformation parameters $\{ a_{13}, \varepsilon_3\}$. 
The reported $17\%$ uncertainty in the observed shadow radius \cite{Psaltis_etal2020} translates into an allowed range 
$ -3.6 \lesssim a_{13} \lesssim 6$ and $  -7 \lesssim \varepsilon_3 \lesssim 12 $ for the deformation parameters. It can be noticed that the two 
parameters are anti-correlated: a positive (negative) $a_{13}$ ($\varepsilon_3$) gives rise to a bigger shadow relative 
to the canonical GR radius $b_{\rm GR}=3\sqrt{3} M$ (and vice versa for the opposite signs). 

This, however, is not the end of the story: the apparent shadow radius is also a function of the geometry of the 
illuminating accretion flow~\cite{Gralla_etal2019}.  Assuming GR gravity, this radius (which represents the peak 
of the emitted flux) is given by $b_{\rm GR}$ when the black hole is  `backlit' from a distant uniform source. The same is true for the more 
astrophysically relevant scenario of a spherically symmetric flow in the vicinity of the black hole \cite{Narayan_etal2019,Gralla_etal2019}. 
In contrast, illumination by a thin accretion disk would lead to a somewhat larger shadow radius $b \approx 6.2M$ \cite{Gralla_etal2019}. 
A more realistic alternative possibility for a system like M87* is that of a geometrically thick/optically thin disk; in such a case the analysis of 
Ref.~\cite{Gralla_etal2019} suggests a shadow radius of $b \approx 5.8M$ that lies between the two previous values. 
The matter-induced deviation of the apparent shadow radius from the mathematical value $b_{\rm GR}$ 
has been the subject of more sophisticated modelling in \cite{Bronzwaer:2020vix}. According to this recent work the resulting `error' 
in the radius is $\approx 5\%$ which translates to $b\approx 5.5M$.

The uncertainty in $b$ caused by the unknown accretion physics of M87* is shown in Fig.~\ref{fig:bJ_shadow} as a grey band and has some 
clear implications for the earlier constraints on the deviation from GR. A large portion of the parameter space previously associated 
with an enlarged  black hole shadow as a result of deviations from GR is now occupied by the more prosaic accretion physics `error' \cite{Gralla2021}. 
Only the spacetime deviations for which $b < b_{\rm GR}$ can be cleanly probed by the shadow measurement. Indeed, and given that 
it marks the peak of the geodesic potential, $b_{\rm GR}$ is the absolute \emph{minimum} shadow radius irrespectively of the accretion 
physics details.

The situation could  deteriorate further if both deformation parameters are non-vanishing \cite{Volkel:2020xlc,Psaltis_etal2020}. 
As a consequence of their anti-correlation, the shadow of a black hole with $\alpha_{13} \sim \varepsilon_3$ could lie significantly closer to 
$b_{\rm GR}$ for the same degree of deformation. This is exemplified by the dashed curve in Fig.~\ref{fig:bJ_shadow} which shows $b$ for 
$\alpha_{13} = 1.2\, \varepsilon_3$. It is clear that in such a case most of the deviation away from $b_{\rm GR}$ overlaps with the accretion 
physics error and the constraints on $\alpha_{13}, \varepsilon_3$ are far less reliable. Of course, we would have drawn the exact opposite 
conclusion if $\alpha_{13} \sim -\varepsilon_3$.

The upshot of this discussion is that the quality of black hole shadows as probes of GR gravity could be diluted by the system's matter degrees 
of freedom and by possible (anti)correlations between the non-GR parameters of the metric. An additional complication lies in the shadow's
shape itself; nearly circular shadows are relatively ubiquitous in non-Kerr spacetimes, resembling the shape of a Kerr shadow for most  
of the allowed spin range. As a case in point, consider the parametrised metric of Carson \& Yagi \cite{Carson_Yagi2020} which is an extension 
of the J-metric and represents the most general family of separable and asymptotically flat spacetimes. This is a metric that can be mapped on black 
hole solutions originating from various modified theories of gravity; it too can lead to Kerr-like quasi-circular shadows for a wide range of its deformation 
parameters, especially when the spin is not too high.


\section{The importance of non-GR coupling parameters} 
In a sense, our discussion so far was just the tip of the proverbial iceberg because even if we were to put aside the complications 
related to the physics of the accretion flow and the commonness of quasi-circular shadows, we would still have to face a much more 
serious problem related to the mass dependence of the non-Kerr deformation parameters. The key issue here is to understand to what 
extent the constraints placed on these parameters by the M87* image are compatible with limits placed by gravitational wave observations 
of merging black holes \cite{TheLIGOScientific:2016src,LIGOScientific:2019fpa,Abbott:2020jks} or electromagnetic observations of astrophysical 
black holes in X-ray binaries \cite{Kong:2014wha}. 

The crucial importance of the mass scaling of a gravity theory's non-GR parameters was first discussed in \cite{Barausse:2014tra} and 
was recently emphasized in \cite{Maselli_etal2020} in the context of gravitational waves from extreme mass ratio inspirals (EMRIs) 
in supermassive black holes. In order to understand the impact of the mass, we follow the reasoning of this recent work and consider modified theories 
of gravity, as extensions of GR, described by an action of the following general form,
\be
\cS= S_{\rm GR} (g_{\mu\nu}, \phi) + \alpha\, \cS_{\rm c} (g_{\mu\nu}, \phi) + \cS_{\rm m} (g_{\mu\nu}, \phi, \Psi),
\label{S1}
\ee
where $g_{\mu\nu}$ is the metric, $\phi$ is the scalar field degree of freedom and $\Psi$ stands for the matter fields.
The first term represents the GR part of the action,
\be
 S_{\rm GR} = \frac{1}{16\pi} \int dx^4 \sqrt{-g} \left ( R - \frac{1}{2} \partial_\mu \phi\, \partial^\mu \phi \right ),
\ee
where $R$ is the Ricci scalar and $g$ is the metric determinant. The last term is the matter part of the action and
can be set to zero for the purposes of this paper.
The theory's non-GR physics is encapsulated in the term $ \cS_{\rm c}$ which describes non-minimal couplings between 
$g_{\mu\nu}$ and $\phi$; the factor in front of this term is the theory's coupling constant\footnote{The conclusions of this 
paper should be equally applicable to actions more general than \eqref{S1}, including additional scalar fields and coupling constants.}.

First we need to distinguish between two different scenarios: (i) the Kerr metric is an admitted black hole solution 
of a non-GR theory; (a) this could be the only possibility (as in $f(R)$ gravity~\cite{Psaltis_etal2008}), (b) or just a solution branch among 
other non-Kerr solutions, (ii) the more generic scenario of genuine non-Kerr solutions.
Given that a black hole shadow is essentially the result of photons moving along the geodesics of the 
hole's spacetime, it is clear that the first scenario is associated with shadows identical to the GR Kerr shadows 
and is therefore untestable with this method. The family of theories with non-Kerr solutions (including spin-scalarised 
Kerr black holes \cite{Dima:2020yac,Berti_etal2021,Herdeiro_etal2021}) should generically lead to non-Kerr shadows and 
hereafter we focus on them.

In their vast majority, these theories (and as a consequence their black hole solutions) are endowed with 
a \emph{dimensional} coupling constant that scales as $\alpha \sim M^n$  with $n\geq 1$. 
Examples of  such theories include scalar Gauss-Bonnet gravity, generalised scalar-tensor theories and dynamical 
Chern-Simons gravity \cite{Berti:2015itd}). Black hole metrics in these theories depend on $M$, a spin parameter $a/M$ and 
a \emph{dimensionless} coupling constant
\be
\zeta \equiv \frac{\alpha}{M^n}.
\label{zeta}
\ee
The hole's scalar field is expanded in $\zeta$ around its constant asymptotic value and enters the metric through a dimensionless function
of order unity that multiplies $\zeta$ (e.g., see Ref.~\cite{JulieBerti2019}).
The appearance of $\zeta$ should not come as a surprise since, as we have pointed out, the mass is the system's only available 
dimensional scale. Theory-agnostic deformed Kerr metrics can be typically mapped onto specific `genuine' theories with dimensional constants. 
Examples are provided by the general Carson-Yagi metric mentioned earlier \cite{Carson_Yagi2020}, and the J-metric used in this paper. 
As it turns out, in all cases the deformation parameters are simply related to the coupling constant of the corresponding gravity theory 
(e.g.~\cite{Johannsen:2013pca,Carson_Yagi2020}). For the case of the J-metric we typically have $\alpha_{13} \sim \zeta^k$ with $k\geq 1$, 
and similarly for the other parameters.  

At this point we may return to the analysis of the M87* shadow and examine what are the implications of using non-GR models with
dimensional constants for its description. As we have seen in the previous section the constraint on $\alpha_{13}$, 
when expressed in terms of the coupling constant $\zeta$ of a given theory, amounts to
\be
 |\zeta| \lesssim s ~\Rightarrow~ |\alpha| \lesssim s M^n,
 \label{ineq1}
\ee
where $s \sim 10$ approximately. We imagine that the same non-GR model is also used in the study of the gravitational wave-driven
inspiral and merger of a black hole binary system of typical mass $M_{\rm b}$, resulting in a similar constraint 
$|\zeta_{\rm b} | \lesssim s_{\rm b} $, where
\be
\zeta_{\rm b} = \frac{ \alpha}{M_{\rm b}^n}= \zeta  \left ( \frac{M}{M_{\rm b}} \right )^n.
\ee
Existing limits from astrophysical observations (see e.g. \cite{Blazquez-Salcedo:2016enn, Nair_etal2019}) suggest $s_b \lesssim1$. 
Using this as a fiducial limit for our J-metric model, we find
\be
|\alpha_{13}| \sim  |\zeta_{\rm b}|^k \left ( \frac{M_{\rm b}}{M} \right )^{kn} \lesssim 10^{-8k n}  \left (\frac{M_{\rm b10}}{M_9} \right )^{kn},
\ee 
where the masses have been normalised to their typical values, $M_9 = M/10^9\,M_\odot$, $M_{\rm b 10} = M_{\rm b}/10\,M_\odot$. 
Thus we have shown that $|\alpha_{13} | \ll 1$; as a consequence of the mass scaling of $\zeta$, 
a similar result should hold for the rest of the parameters since all of them are  comparable to $\zeta$.

The same argument can be turned around: a typical deformation $\alpha_{13} \sim  {\cal O} (1)$ coming from the shadow 
of M87* would be stretched by a factor $ \sim (M/M_b)^{kn}$ when the J-metric is used to model the celestial mechanics  of 
a merging  binary system. This would cause an enormous deviation from the GR black hole metric which would have easily 
been seen in the system's GW signal.

The mass-suppression effect could be evaded if a black hole is exactly described by the Kerr metric
within a non-GR theory but could undergo a spin-induced scalarisation above a spin threshold ~\cite{Dima:2020yac}  
(i.e. this is the previously mentioned scenario (ib)). An example is provided by Gauss-Bonnet gravity itself 
which for a vanishing derivative of the scalar field coupling function does indeed admit the Kerr solution. 
Recent work~\cite{Dima:2020yac,Berti_etal2021,Herdeiro_etal2021} suggests that such black holes can become non-Kerr by spontaneous 
scalarisation if they reside in a wedge-shaped region bounded by $a \gtrsim 0.5M$ and a negative coupling 
$\alpha/M^2 \sim -(0.1-10)$ . Although this region represents a small fraction of the parameter space it is possible 
to imagine a scenario in which stellar-mass black holes probed by GW observations lie outside the scalarisation wedge
 (and therefore are Kerr) whereas rapidly spinning supermassive black holes (a viable possibility for M87*)  are scalarised and non-Kerr.

The remarkable conclusion of this section is that one should typically expect (at least for most of the straight forward extensions of GR)
the black hole spacetime of M87* to be described by the Kerr metric to a very high precision, with all non-Kerr deviations suppressed by the system's enormous mass. Once the metric is rendered Kerr for all practical purposes, it follows that all geodesic motion and the shadow itself must necessarily 
be also Kerr~\cite{Barausse:2014tra,Maselli_etal2020}.


\section{Concluding remarks} 
The take home message of this paper is rather clear: the shadow appearing in a black hole image like that of M87* 
could be a viable probe of GR gravity (and more specifically of the Kerr spacetime) but with some important caveats attached. 
This standpoint lies somewhere in between the recent opposing claims made in Refs.~\cite{Psaltis_etal2020,Gralla_etal2019}
but at the same time it extends to a completely orthogonal direction. 

It is certainly true that the shadow radius is primarily a function of the black hole's spacetime but also of the (largely unknown) accretion 
flow physics. However, if GR gravity is assumed, the radius cannot be pushed below $b_{\rm GR}$ and therefore a $b < b_{\rm GR}$ 
ought to be a clean probe of the black hole's spacetime metric (provided it is observationally allowed in the first place). 
Moreover, the constraints placed on the deformation parameters of the non-Kerr model also depend on their actual number 
\cite{Volkel:2020xlc}. 
The quasi-circular shape of the M87* shadow is another complicating factor because similarly shaped shadows 
commonly emerge in non-GR gravity theories and deformed black hole spacetimes alike. The very recent work of 
~\cite{Junior:2021atr} represents a detailed study of the degeneracy between Kerr and non-Kerr black holes in 
the strict sense of exactly matching shadows; of course a more empirical approach is also possible 
taking into consideration that observational errors can easily accommodate a small mismatch between shadows. 

Our discussion of non-GR theories with coupling constants has revealed a perhaps unexpected dichotomy: the shadow cast by
a supermassive black hole is intrinsically insensitive to deviations from Kerr when the underlying gravity theory contains dimensional
coupling constants (in which case Eq.~\eqref{zeta} shows that the deviation parameters are mass-suppressed). The majority of
known extensions of GR do indeed fall into this category. 
Nevertheless, there exist notable exceptions like Einstein-\ae ther theory~\cite{Jacobson_Mattingly2001} where the non-GR 
parameters are dimensionless quantities. The non-Kerr character of black holes in such theories is equally prominent regardless 
of their mass, and therefore their shadow \emph{could} be used as a test of GR (see \cite{Khodadi:2020gns} for a calculation 
along these lines in the context of  Einstein-\ae ther theory) albeit subject to the influence of the factors discussed in this paper. 
Even within the class of theories with dimensional coupling constants there are two mechanisms that 
could invalidate the mass-suppression effect but, as we argue, these are likely to be the exceptions to the general rule.
The first (and most interesting) one is the spin-induced scalarisation discussed in the previous section. The second 
mechanism is that of screening; many theories rely on screening mechanisms in order to survive as viable extensions of 
GR `across the spectrum' from solar system tests and compact object binaries out to cosmological scales (for a review see~\cite{Joyce:2014kja}). 
However, most of these mechanisms (such as chameleons and symmetrons) act through the coupling of scalar fields with matter and as a result
black holes can become impervious to them in vacuum models (with  $\cS_{\rm m} =0$).
Screening could also operate via the presence of non-linear interactions in the Lagrangian (as in the commonly used Vainshtein 
mechanism~\cite{Babichev:2013usa,Joyce:2014kja}) but in that scenario it would take some fine-tuning to screen the solar-mass black holes probed 
by GW observations while leaving supermassive black holes unscreened.

The suppression of the non-GR coupling constants in the metric of massive systems is likely to have much wider 
repercussions than what discussed here. Apart from its impact on EMRIs~\cite{Maselli_etal2020} 
and black hole spectroscopy by LISA~\cite{Maselli_etal2020b}, 
we would also expect that electromagnetic radiation (such as the observed X-ray iron lines, continuum emission, or quasi-periodic oscillations) 
from accretion disks in active galactic nuclei, being emitted or reflected by matter moving on geodesics, to be almost completely oblivious to 
deviations from GR gravity \cite{Liu:2014awa,Tripathi:2019bya}. The same should be true for any astrometric observations 
of bodies in orbit around SgrA* supermassive black hole \cite{Psaltis:2015uza}.  We plan to explore some of these issues in the near future.

%
\acknowledgments

We thank Hector Silva, Emanuele Berti and Thomas Sotiriou for useful discussions and helpful feedback.
KG acknowledges support from a  Fundaci\'on Seneca (Region de Murcia) grant No. 20949/PI/18. 


\bibliography{biblio.bib}

\end{document}